\documentclass[utf8]{frontiersSCNS} 

\usepackage{url,hyperref,microtype,subcaption}
\usepackage[onehalfspacing]{setspace}

\def\keyFont{\fontsize{8}{11}\helveticabold }
\def\firstAuthorLast{Orlosky {et~al.}} 
\def\Authors{Jason Orlosky\,$^{1,9,*}$, Misha Sra\,$^{2}$, Kenan Bekta\c{s}\,$^{3}$, Huaishu Peng\,$^{4}$,
Jeeeun Kim\,$^{5}$, Nataliya Kos'myna\,$^{6}$, Tobias Hollerer\,$^{2}$, Anthony Steed\,$^{7}$, Kiyoshi Kiyokawa\,$^{8}$, and Kaan Ak\c{s}it\,$^{7}$}

\def\Address{
$^{1}$Osaka University, Cybermedia Center, Osaka, Japan \\
$^{2}$University of California, Santa Barbara, Computer Science Department, Santa Barbara, California, United States of America \\
$^{3}$University of St. Gallen, Computer Science Department, St. Gallen, Switzerland\\
$^{4}$University of Maryland, Computer Science Department, College Park, Maryland, United States of America\\ 
$^{5}$Texas A\&M University, Computer Science Department, College Station, Texas, United States of America \\ 
$^{6}$Massachusetts Institute of Technology, Computer Science Department, Cambridge, Massachusetts, United States of America \\
$^{7}$University College London, Computer Science Department, London, United Kingdom \\ 
$^{8}$Nara Institute of Science and Technology, Division of Information Science, Nara, Japan \\
$^{9}$Augusta University, School of Computer and Cyber Sciences, Augusta, Georgia, United States of America
}
\def\corrAuthor{Corresponding Author}

\def\corrEmail{email@uni.edu}

\begin{document}
\onecolumn
\firstpage{1}

\title[Telelife]{Telelife: The Future of Remote Living} 

\author[\firstAuthorLast ]{\Authors} 
\address{} 
\correspondance{} 

\extraAuth{}

\maketitle

\begin{abstract}

\section{}
In recent years, everyday activities such as work and socialization have steadily shifted to more remote and virtual settings. 
With the COVID-19 pandemic, the switch from physical to virtual has been accelerated, which has substantially affected various aspects of our lives, including business, education, commerce, healthcare, and personal life. 
This rapid and large-scale switch from in-person to remote interactions has revealed that our current technologies lack functionality and are limited in their ability to recreate interpersonal interactions.
To help address these limitations in the future, we introduce ``Telelife,'' a vision for the near future that depicts the potential means to improve remote living better aligned with how we interact, live and work in the physical world.  
Telelife encompasses novel synergies of technologies and concepts such as digital twins, virtual prototyping, and attention and context-aware user interfaces with innovative hardware that can support ultrarealistic graphics, user state detection, and more. 
These ideas will guide the transformation of our daily lives and routines soon, targeting the year 2035.
In addition, we identify opportunities across high-impact applications in domains related to this vision of Telelife. 
Along with a recent survey of relevant fields such as human-computer interaction, pervasive computing, and virtual reality, the directions outlined in this paper will guide future research on remote living.

\tiny
 \keyFont{ \section{Keywords:} Telelife, Telepresence, Telework, Remote Living, Digital transformation} 
\end{abstract}

\begin{figure}[ht]
\begin{center}
\includegraphics[width=17cm]{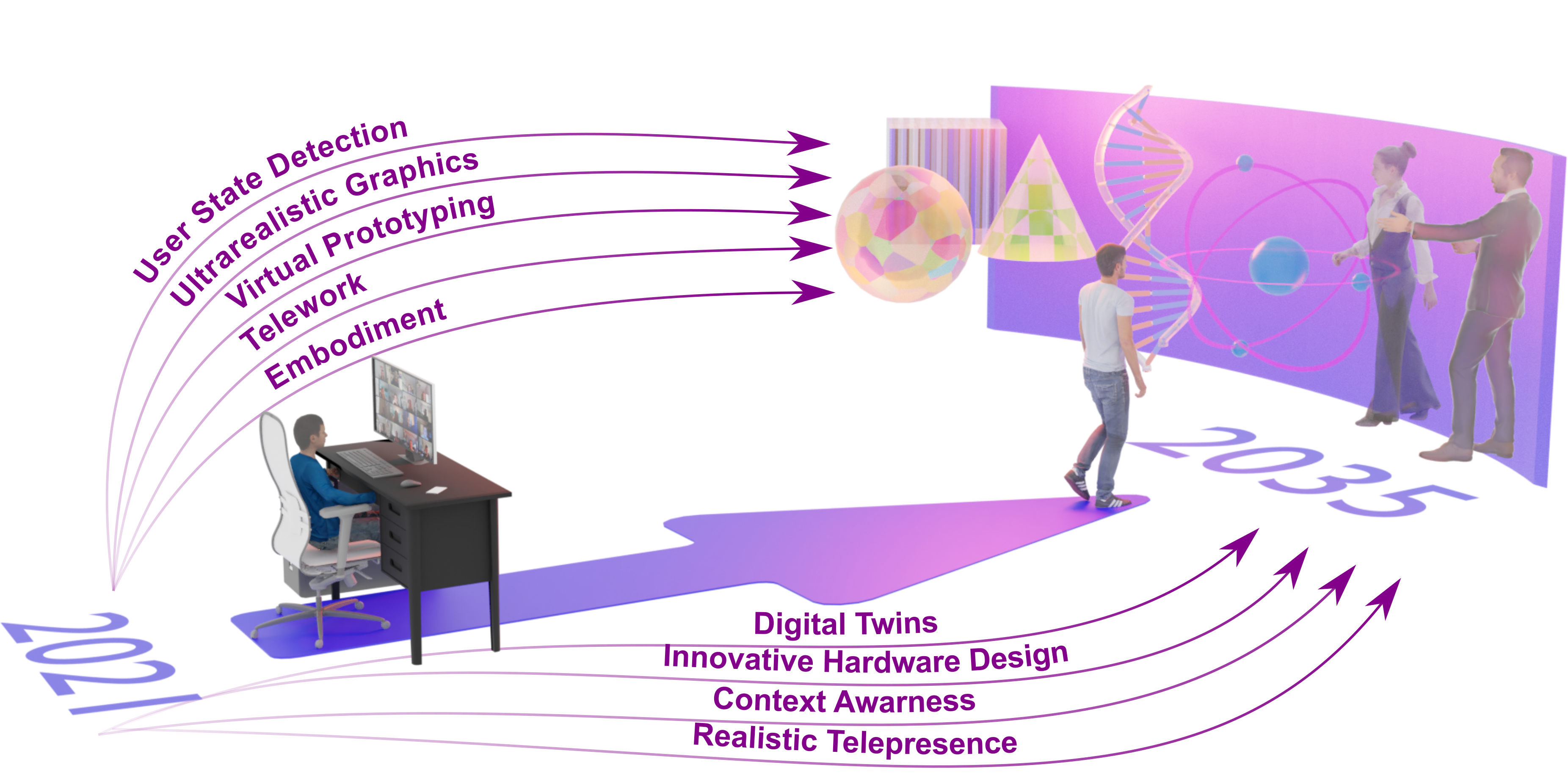}
\end{center}
\caption{Telelife, the future of remote living. In 2021, we are still bound by our technology and devices, practically locked in front of a stationary display and operating remotely in an unnatural and often exhausting way.
Telelife provides a vision of the future that looks towards 2035, where a majority of our interactions and our daily routines will be supported with remote technologies that are aligned with human nature.
Telelife integrates learning, exploration, interaction problem solving, and even gaming into a virtual ecosystem in which users will have experiences that are closely integrated into their lifestyles.
To achieve this future, breakthroughs will be needed along various research topics such as user state detection, ultrarealistic graphics,  virtual prototyping, telework, embodiment, innovative hardware design, digital twins, context awareness and realistic telepresence.}
\label{fig:teaser}
\end{figure}

\section{Introduction}
In 2020, the COVID-19 pandemic forced us to change the way we live. A major part of our daily lives became remote~\cite{steed2020evaluating}, and this change represented a major shift in how we work and socialize.  
For example, society realized that remote work can promote greater time flexibility, collaboration, and efficiency in working hours. 
Consequently, the ideas of telepresence~\cite{held1992telepresence} and telework~\cite{huws1990telework} regained momentum, presenting themselves as a meaningful way to overcome the limitations of remote work centered around video conferencing.

However, the promises of telepresence and remote interaction have yet to be achieved. 
The sudden, forced shift to remote operations without adequate preparation has produced an immediate need for new tools, research, and policies.
Commercially available tools for remote operations inherently have many significant challenges, such as lack of 3D visuals and interactions.
These existing tools in the current transition towards remote technologies are very much in their infancy.
With the recent changes, new issues related to ethics, social behavior, and accessibility have also arisen and are starting to be discussed among practitioners and researchers.
For example, many users lost the close interpersonal feedback that they typically received through body language and physical interaction with others. 
The loss of interpersonal feedback is partly due to the lack of high fidelity visual feedback between themselves and their counterpart, sometimes resulting in a loss of attention and context-awareness critical for collaborative work and sometimes in an isolated or overwhelmed feeling in online settings.

This paper provides a review of technologies encompassed in an ecosystem that we call ``Telelife.'' This represents an assembly of remote technologies and new innovative ways to enhance the way we live our intellectual, professional, community and personal lives in a remote context. 
This draws from prior work in telepresence, telework, and supporting technologies such as augmented reality (AR)~\cite{azuma1997survey}, virtual reality (VR)~\cite{wang2002definition}, pervasive computing, and relevant sub-brabnches.
Telelife unifies these works into a single space, and we highlight recent state-of-the-art research that has brought society closer to remote living. 
While existing concepts often focus on the design of enabling technologies for remote collaboration or specific applications, telelife offers a broader perspective that holistically addresses our needs as humans. 
As such, we revisit the need for new tools and concepts in remote operations, emphasising the technologies necessary to support remote living in the next era.

Moreover, numerous issues related to user adoption of 3D technologies remain as major bottlenecks due to technical limitations, such as the lack of small form-factor near-eye displays (NEDs) or effective sensing devices.
Other needs include the transmission and acquisition of realistic 3D graphics, which are significant challenges in  computing and graphics.
The greatest challenge for the realization of telelife is to enable seamlessly integrated technologies in people's lives while providing experiences in line with human nature and psychology.

The key contributions of this work include the identification of potential technologies that will help enable telelife and a synthesis of the highest impact research from various fields to help define the future of an increasingly remote society.
Another goal of this synthesis is to help researchers find opportunities for growth in relevant fields and guide new research directions within the emerging telelife ecosystem. 
More specifically, our contributions, which also represent the layout of the paper, are as follows:

\vspace{-0.1cm}
\begin{itemize}
\itemsep 0em 
\item Unified Research Vision of Telelife. We help shape the vision for the near future by merging and unifying the highest-impact use cases and identifying interactions, technologies, and new paradigms within those areas. 
We name this agenda Telelife, a synthesis of discrete research visions.
\item A day in 2035. We introduce a series of short speculations that exemplify a day in the telelife of 2035, in various aspects of life.
\item Challenges in Telelife. This section highlights challenges with remote interfaces, including smart homes, learning, collaborating, privacy, security, accessibility, adoption, and ethics.
\item Innovative new research directions. We provide guidance for researchers from a diverse set of fields to define the next steps in telelife related research.
\item Grand technical challenges. We provide a dense summary of the technical challenges associated with realizing this research vision and conclude with a discussion of new telelife interactions.
\end{itemize}

\section{The Vision of Telelife}
For many years, researchers have dreamed of changing what we know as teleconferencing into telepresence. 
While this goal has merit, replicating a remote environment such as a classroom or office in its entirety still poses significant challenges. 
In contrast to the vision of telepresence, our survey named telelife will not only match the benefits of in-person meetings but will wildly exceed them.

Moreover, social media has had the unfortunate side-effect of reducing face-to-face meetings and interpersonal interaction. 
AR, VR, and related fields are currently in a unique position to ``re-humanize" digital interactions between people. 
Though it has become possible for individuals to improve dialogue through messaging, images, and live video, we have yet to replicate the benefits of minute facial movements, eye contact, body language, emotion, and expression that make up a large portion of interpersonal communication. 

We distil telelife down into four core areas: intellectual, professional, community, and personal life.
Humans have a natural tendency to explore, interact, and socialize in these contexts in 3D. 
What is essential is to focus on the right combination of research topics that will result in a state where the advantages of teleinteraction outweigh the disadvantages. 
It is not to say that 2D data representations such as spreadsheets or charts will become obsolete. 
However, telelife will afford users a much richer experience and propel us past the cubicle, classroom, and desk paradigms pervasive in today's world. 
Especially for small group interactions and creative work, 3D representations will provide the opportunity to replicate and exceed the experience of round-table discussions in physical space.

Additionally, just like collaboration platforms have ``channels,'' Telelife will have its new 3D multitasking and messaging systems class. 
These will become an essential part of the ecosystem that takes us into the next generation of 3D interfaces. 
More importantly, this ecosystem will need to ensure that new interactions are not invasive, forced, detrimental to human productivity, or dehumanizing. 
Privacy, access control, and controllable interaction will be essential, and great care should be taken to ensure that new research adapts to the telelife paradigm without hindering interpersonal interaction.

The ideas discussed in this paper build upon the culmination of previous work in fields such as AR, VR, telepresence, communications research, and pervasive computing. 
In coming up with the vision of telelife, we cherry-picked the most helpful and highest impact concepts from these fields to design the tele-ecosystem of the future. 
For example, while pervasive AR~\cite{grubertetal2017} is embodied by continuous or always-on AR, telelife is focused on the integration of information as a part of the user's life experiences, socially, emotionally, and intellectually. 
Table~\ref{table:telelife_comparison} provides a summary of the commonalities and differences between telelife and other relevant fields.

\begin{table*}[!htp]
\scriptsize
\caption{Telelife distinguish itself from pervasive augmented reality, telepresence or virtual reality by offiering a unique blend of remote technologies that offers new innovative ways to enhance the way we live our intellectual, professional, community and personal lives.}
\resizebox{\textwidth}{!}{
\begin{tabular}{lrrrrr}\toprule
&Pervasive Augmented Reality &Telelife &Telepresence &Virtual Reality \\\midrule
Use &Continous &User aware &Sporadic &Sporadic \\
Control &Context controlled &Both user and context controlled &User controlled &User controlled \\
Interaction with physical environment &Visual (observing) &Bidirectional (physically) &Bidirectional &Isolated \\
Context of use &Context aware &Seamless &Specific &Specific \\
Ease of adoption &Learnable &Inherent to human nature &Learnable &Application based \\
Ease of use &Artificial &Inherent to human nature &Inherent to human nature &Close to human nature \\
\end{tabular}
}
\label{table:telelife_comparison}
\end{table*}

In short, telelife blends all aspects of the virtual spectrum into real life, focusing on the ecosystem in which users interact rather than focus on specific applications.
It is about the role of telepresence and advanced displays in the entirety of our daily lives, and it is not just constrained to specific spaces or roles. This paper functions as a guide for researchers and industry to focus efforts on developing 3D interactions, reconstruction techniques, and telepresence technologies that will propel society into a new age of remote living.
\section{A day in 2035}
September 5th, 2035, Wednesday. 
Mike wakes up early, hoping to do something special for his son Jimmy's tenth birthday.
While still in bed, he blinks to wake up his telelife twin and asks him to prepare breakfast. 
Mike does not need to go to the kitchen personally. Instead, all available ingredients in the fridge are listed virtually in front of him with recommended recipes. 
His smart kitchen equipment can automatically prepare the food while Mike monitors the cooking progress through his twin's first-person view, and he chimes in to add a final touch of ground black pepper to his eggs.

It is now 8:30 am. 
Jimmy has finished breakfast and is ready for his school day. 
Every Wednesday is history day at teleschool, and Jimmy does not need to travel there physically.
Instead, Jimmy switches the telechannel to "school" to meet with his teacher and friends in a fully immersive virtual space automatically customized for his physical room.
As usual, kids in the classroom share 3D emojis that are shared only between their personal views so as not to interrupt others.
The first class of the day is history.
In this virtual space, Jimmy is now sitting in an old-fashioned classroom with his classmates, each of whom designed their virtual uniform as homework for the historic visit.
The teacher's avatar takes the shape of an old-fashioned, early 21st-century citizen to create the right ambience for the topic.
"Today's class is about global pandemics. We are going to experience the Spanish flu of 1918 and the COVID-19 pandemic of 2020 first-hand. 
To better illustrate, I am first setting our digital environment time slider to the year 2020. 
Let us zoom into our current location to observe what was taking place in September."
Jimmy begins his journey through a day in 2020 to feel the impact of the crisis and learn about the need for global preparedness.

While Jimmy is learning history at teleschool, Mike quickly browses today's news before setting off to work.
The newspaper has long been outdated in 2035. 
Instead, the latest 3D reconstructed news scene is rendered as a small-scale model that travels along with Mike as he moves.
He glances over today's stock prices, each showing historical balance sheet data rendered as a simple graph that an air-swipe of his finger can reorganize.
At this moment, Mike is reading about a company specialized in personal garment manufacturing while feeling the fine texture via his virtual-touch interface.
While the 3D data shows cash flows and reduced debt, a recent fashion show sponsored by the company floats alongside.
Within a short time, multi-channel and multi-model information is organized to help Mike decide whether he wants to buy a piece of the business.

Mike is now ready for work. 
Although he does not need to go to the office physically on a daily basis, Mike decides to go today.
He has been working on a big architecture project in collaboration with a client in Asia. 
Today, Mike will demonstrate his model to his client using remote tele-fabrication.
In his office, Mike first turns off the virtual sun since it is nighttime in Asia. 
He then logs into the digital meeting space with his clients. 
Mike and his clients are now co-located in the same virtual office in a split second, equipped with a tele-fabrication machine.
Back in physical space, Mike's display can overlay authentic 3D visuals anywhere and at any size. 
His room is also fitted with a high-fidelity, mid-air haptic feedback device synced to a display with less than a nanosecond of latency. 
In the physical space in Asia, his clients stand in front of a rapid prototyping machine that responds to Mike's remote actions in real-time.
As Mike starts to explain the design concept while replaying his design process in the digital space, the 3D physical model is seamlessly built-in reality in front of his clients.
His clients all keep their own small, physical copy of the model to take home and interact with later.

It has been a long day, and Mike's telelife twin reminds him of Jimmy's birthday.
Mike and his telelife twin have been together since 2023. 
His twin has been his most immense emotional support, especially after Jimmy's mom passed away three years ago.
After a long but productive day of telework, Mike leaves his office, sits in his self-driving car and browses Jimmy's 3D childhood photos and asks his telelife twin for a suggestion on cake design.
Twin-Mike suggests the one with Jimmy standing and smiling next to a giant furry monster from tele-playground. 
That was a somewhat lower resolution 3D-voxel map taken when Jimmy was 3 when voxel-based imaging was still in its infancy. 
Fortunately, Mike's digital twin has the latest AI retouching upgrades, so he produces a superresolution version for Mike to send to a local bakery for a customized birthday cake.
Mike's twin's mainframe runs on the most recent photonic computer that runs on renewable energy resources. 
Though inexpensive twin-service is available, Mike maintains his twin's mainframe himself.
It is set to auto-update with the help of a local chapter of the combined intelligence agency, in which he is also an active member.

It is now 6 pm. 
Mike is home and celebrating Jimmy's 10th birthday. 
Mike invited all his friends to Jimmy's tele-birthday and asked everyone to come up with their digital costume.
They all talk, chat, draw and play around a large, interactive virtual table, and for many days following the event, Jimmy reminisces about the abounding laughter.

Fast forward to 2065. 
Jimmy has just turned 40, and he feels an internal warmth as he replays the moment from his 10th birthday in his mind using his direct brain-computer interface (BCI).
To experience the old days, he decides to deactivate his BCI and enjoys his vacation on a sunny beach with his loved ones face-to-face for a time.
"So much has changed," Jimmy says. So many discussions have taken place along the way to adopt these technologies, and so many political conflicts have been resolved by bridging the world through telelife.
In the end, Jimmy is glad that we have been able to re-humanize the way people live and communicate in today's teleworld. 
\section{Challenges in Telelife}
Functional telelife, as presented in the scenario above, faces several unique challenges, both technical and practical.  
While several of the technologies discussed are already available in some form, smooth integration between technologies, devices, and life presents many hurdles. Though much research has been done to address communication, interaction, display form-factors, participant visualization, and user feedback, much of the work done up to now has not yet become standard practice. 

This section discusses some of the main challenges of achieving these in practice: interactions with intelligent homes, classroom and remote learning, collaboration with remote colleagues, and broader challenges related to privacy, security, ethics, accessibility, and adoption.  

\subsection{Smart Homes} 
Smart home technologies constitute digitally controllable systems like lighting, heating and cooling, and Internet of Things (IoT) devices like kitchen appliances, robotic vacuum cleaners, and lawnmowers.
For a smart home to function as envisioned, it must reliably interpret the inhabitants' movements and expectations, distinguish usual patterns from exceptions, correctly interpret inhabitant physiological signals and respond accordingly. 
The essence of a smart home is the inter-connectedness of different automated systems. 
However, fractured ecosystems, limited interoperability of heterogeneous technologies, and the lack of regulation and standards~\cite{stojkoska2017review} remain significant obstacles to the realization of this goal.

In addition, the security and privacy concerns of the inhabitants may lower the acceptance of such technologies~\cite{brushetal2011}.
These challenges need to be addressed to support seamless integration of the smart home, IoT devices, and entities like Mike's digital twin, as envisioned in the scenario above. 

\subsection{Learning} 
The introduction of rapid prototyping and immersive technologies like AR and VR have enabled newer forms of learning yet. 
High-quality interaction with the learning environment has been shown to improve learning~\cite{dalgarno2010learning}, and AR has been shown to outperform paper-based and digital methods of instruction delivery in terms of accuracy and time~\cite{bhattacharya2019augmented}. 
However, these new learning styles are not widespread because authoring content that is interactive, immersive, and educational is a complex process that lacks standardization and requires technical skills to create.
While the use of AR-enhanced books is a powerful educational tool \cite{billinghurst2002augmented}, designing technologies for children, for example, requires taking into account their developmental abilities. However, there is relatively little work in that field.

To realize the immersive remote learning scenario, we need to start with the senses. 
Sensory modalities including vision, audio, haptics, smell, taste, proprioception, and vestibular sense present several actively explored challenges in ongoing research. 
For example, a display's field of view and resolution should match the human visual perception. We need improved spatialization and sound synthesis for audio and a wider variety of sensations for haptics. 
All these need to be integrated into a single system that enables the user to engage with digital content and remotely locate users as seamlessly as possible with physical objects and people in the same room. 

\subsection{Collaborating}
Digital collaborative spaces can help realize mediated social experiences where distance disappears and interact as richly with remotely located users as those in the same room.
Collaborative spaces can be both 2D and 3D and take the form of video conferencing tools (e.g., Zoom, Gather Town), multiplayer video games (e.g., Fortnite), virtual office spaces (e.g., Arthur, Spatial) or virtual communities (e.g., VRChat). 
Collaborative virtual environments involve the use of networked systems to support group work~\cite{dourish1996your}. 

The key concept behind these spaces is that each occupant is represented in a graphical form and can interact with the environment and the other occupants~\cite{churchill1998collaborative}.

An ideal collaborative space would track all user movements, include real-time facial capture for mapping speech, facial expressions and eye gaze for realistic interactions, and provide a space where a user's can move and engage with each other as if they were in the same room. 
It is difficult because either the user's need to wear much equipment to enable tracking or the physical space needs to be instrumented and both have their challenges.
One big challenge in immersive collaborative environments is the lack of input techniques that are as efficient as the keyboard and mouse. 
Due to the limited number of input techniques, the recording, tagging, transcribing, capturing, highlighting and sharing content that is becoming popular with videoconferencing tools are potential research areas.

\subsection{Privacy and Security}  
As with other networked applications, privacy and security are paramount for remote interaction~\cite{steed2009networked}.
With the increase in online interactions involving video or VR, we have seen incidents of Zoom bombing, fake identities, conference call hijacking, deepfakes, and other problems leading to unwanted disruption and distress.

There are serious repercussions for privacy and security as some AR and VR companies start to connect how we move and interact in the virtual and physical worlds with how we think and feel. 
While information about the physical environment is needed to provide the immersive experience or keep the user safe, aggregating that data with biometric and other data further exposes users. 

Because of the data flow in an AR/VR system, what is "seen" may not be intended by the developer since the rendered view may be vulnerable to alteration or injection.
Each step of data flow, including detection, transformation, and rendering (i.e. inputs, operations, and outputs), should be protected, but each step requires different types of protections
at the hardware, software, and communication level~\cite{happa2019cyber}.

Researchers have discovered vulnerabilities associated with some platforms (e.g., HTC Vive and the Oculus Rift) related to being able to change what the user is viewing, and also related to data flowing to/from the headsets \cite{yarramreddy_forensic_2018}.
Changing what the user is viewing could lead to physical harm if safe movement boundaries are altered.
By examining stored data on the VR platform, information about sites visited, time and date information, and user logs can all be recovered, similar to how file forensics of other system files can be recovered.
Informing the user of safe or secure content in AR/VR is desirable, for example, by using a lock icon like that present in the address bar when visiting secure websites. 

Physical characteristics of a user, such as head and hand position and orientation, can be used to identify users, leading to loss of privacy successfully. 
\cite{miller_personal_2020} claims that 95\% of users can be identified using this type of movement tracking information after only 5 minutes of gathered data for training the machine learning identification system. Since movement information is normally captured for research purposes, standard data anonymization techniques will not prevent user identity from being determined from physical characteristics.

\subsection{Accessibility and Adoption} 
According to the CDC, 1 in 4 US adults has some type of disability~\footnote{\url{https://www.cdc.gov/media/releases/2018/p0816-disability.html}}). 
With the influence of COVID-19 on lifestyles, video conferencing has become a prime way to work, attend school, take care of fitness or socialize.
Simultaneously, interest in video games has risen, offering an escape from the pandemic but also providing a new way to connect with family, friends and with colleagues~\footnote{\url{https://www.nytimes.com/2020/07/31/business/video-game-meetings.html}}. 
Video games are also increasingly being used for non-entertainment purposes such as education~\cite{gee2003video} and rehabilitation~\cite{howcroft2012active}. 

However, those who benefit significantly from remote and interactive technologies can often be shadowed by hardware and software accessibility limitations. 
Research has explored eye gaze~\cite{liu2020orthogaze}, speech input, or mouth-based input devices with varying levels of success to enable users to interact with their computers, especially those with motor impairments due to amyotrophic lateral sclerosis (ALS), muscular dystrophy or cerebral palsy. 
While gaming systems~\footnote{\url{https://www.xbox.com/en-US/community/for-everyone/accessibility}} and newer video games~\footnote{\url{https://steamcommunity.com/app/620980/discussions/1/1696046342864856156/}} have come a long way in providing accessibility options, the limitations are mostly related to input devices. 
Newer technologies like gaze-based methods or brain-to-computer interfaces (BCIs) have the potential to empower users. 
While there is ongoing research in these areas, much work needs to be done to make the devices as efficient and readily available as a keyboard and mouse.

\subsection{Ethics} 
Historian Melvin Kranzberg's first law of technology states, "technology is neither good nor bad; nor is it neutral." \cite{kranzberg1986technology}. 
The societal and personal implications of any technology ultimately depend on its use.
For example, content moderation swiftly became a primary research topic in social computing, covering a variety of themes from 
toolkits to prevent email harassment~\cite{squadbox},
to studying unpleasant social behaviours such as cyberbullying in online games~\cite{GameCyberbullying}
Social media platforms and remote collaboration tools that allow the use of text, image and 3D content sharing are also seeing increasing amounts of copyright violations, harassment and hate speech \cite{mahar2018squadbox}.
More recently, with the growing prevalence of AI technologies, the phrase "seeing is believing" is becoming less accurate, as videos, images and voice media are easily and effectively manipulated. 
The risks grow manifold as we move from the screen and begin to inhabit 3D spaces where an avatar may look and talk like someone you know but is a 3D deepfake, generated for malicious purposes.
Moderating content and regulating information policies would be essential to prevent damaging consequences in the future fully connected telesociety.
\begin{figure*}[ht!]
    \includegraphics[width=\linewidth]{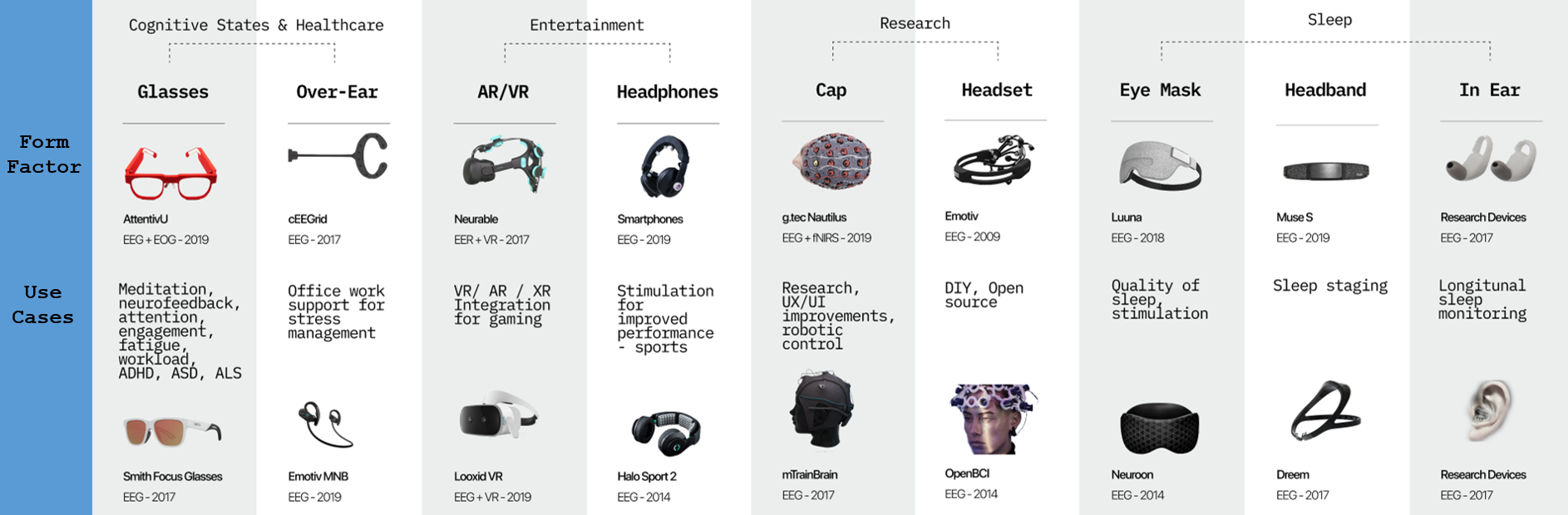}
    \caption{The state-of-the-art in Brain-Computer Interface (BCI) systems discussed in Section~\ref{sec:innovative_hardware}. The figure represents current most common form-factors and use cases of BCI technology for the next 10 years.}
    \label{fig:bci_devices}
\end{figure*}

\section{Innovative New Research Directions}
The potential exists for a multitude of new experiences and interactions in our telelife vision. 
In this section, we lay out areas of research that are ripe for innovation and that will potentially have a high impact on the efficiency, engagement, and progress of remote living in the future.

\subsection{Innovative Hardware Design}
\label{sec:innovative_hardware}
New hardware devices can pave the way to improve realism in our future experiences with telelife technologies while also providing a way to enhance software interfaces with a rich set of user data.
In this section, we focus on innovative hardware designs for generating visuals and capturing a user's attention and state.
It should be noted that future-looking hardware designs go beyond the ones that we cover in this section as a perfect telelife experience would also require a multimodal hardware to support haptic feedback~\cite{krogmeier2019human}, realistic audio~\cite{zotter2019ambisonics} and replication of olfactory experiences~\cite{matsukura2013smelling}.

\subsubsection{Hardware for Sensing User State}
\paragraph{Brain-computer interfaces.}
There are different ways to measure user’s cognitive and mental states like attention, engagement or performance. A lot of studies and therapies turn towards more objective ways of measurement, in particular, physiological sensing. 
Typically, physiological sensing requires the user to wear a set of electrodes and devices that enable real-time monitoring of the user’s physiological and cognitive states. 
Examples include, but are not limited to the use of Electroencephalography (EEG) – non-invasive measurement of brain activity of the user; Electrooculography (EOG) – non-invasive measurement of eye movements of the user; Electrocardiogram (ECG) – non-invasive measurement of heart rate of the user or Electrodermal Activity (EDA) – non-invasive measurement of skin conductance. 
EEG and EOG can provide data relevant to information about attention and engagement of the user~\cite{10.1145/3342197.3344516}. 
Electrocardiogram (ECG) and Electrodermal Activity (EDA) are often used to understand emotional arousal and to identify the magnitude of the emotional response ~\cite{10.1145/2632048.2636065}.
Electromyography (EMG) provides data on facial expressions linked to positive or negative valence~\cite{SATO200870}.

\begin{figure}[!ht]
    \begin{center}
    \includegraphics[width=0.6\linewidth]{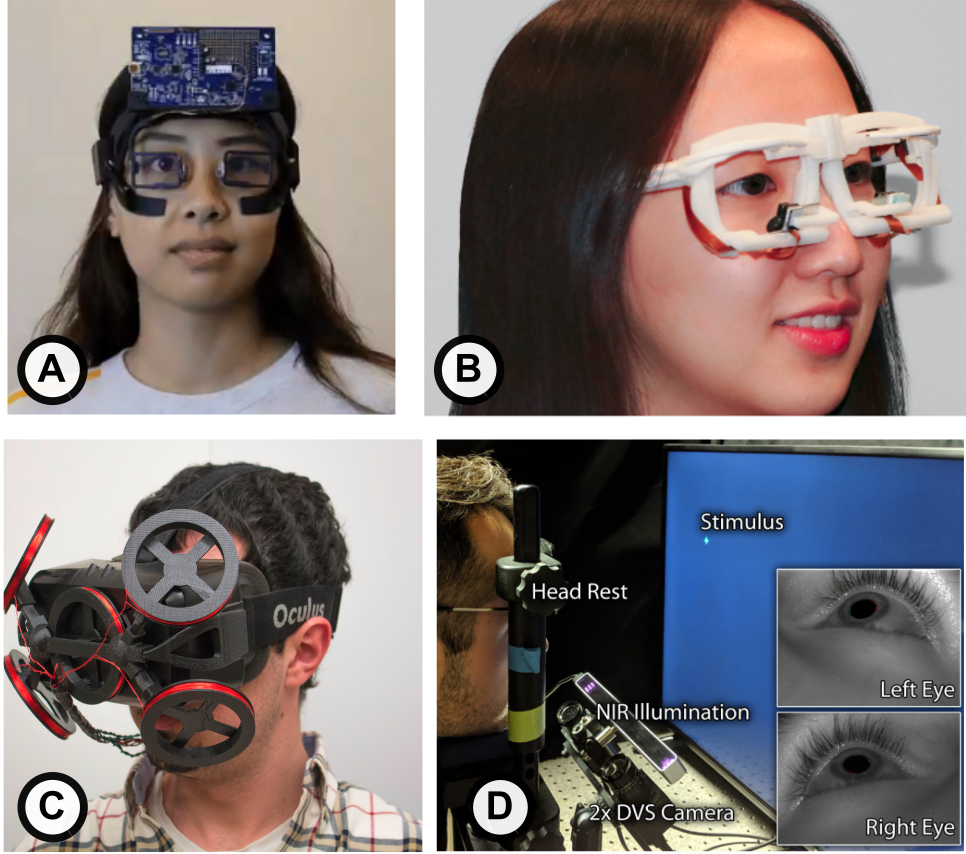}
    \end{center}
    \caption{State-of-the-art in the next generation eye and gaze tracking hardware discussed in Section~\ref{sec:innovative_hardware}: (a) Photodiode and LED based sensing~\cite{li2020optical}, (b) Camera based sensing~\cite{lu2020improved}, (c) Scleral-coil based sensing~\cite{whitmire2016eyecontact}, and (d) Event camera based sensing~\cite{angelopoulos2020event}. Though scleral-coil offers an invasive approach that is against our non-invasive telelife approach, it can potentially be a good match for contact-lens augmented reality near-eye displays at a possible future.}
    \label{fig:eye_gaze_trackers}
\end{figure}

Research in the field of BCI in particular, and physiological sensing in general, has been gaining momentum in the last 10 years, with the systems being used in rehabilitation~\cite{VANDOKKUM20153},
accessing mental states of the user~\cite{10.1145/2556288.2557230} and entertainment~\cite{10.1007/978-3-319-22701-6_37}. However, as of today, these systems still remain expensive, bulky and uncomfortable as gel has to be applied to the electrodes, devices are wired and data is prone to classification errors due to the noisy nature of the signal. 
Thus, a lot of BCI systems nowadays are still being used in association with other input modalities like gaze trackers~\cite{10.1007/978-3-319-22701-6_37}, or VR and AR NEDs~\cite{8329668}.
We believe that this trend will largely expand in the future, and in the next 10 years we will witness more NEDs with integrated hardware to measure brain activity. 
Beyond NEDs, neuromonitoring and neuroenhancing of one's wellbeing is expected to expand in the workplace for managing stress, meditating and measuring attention.
\vspace{0.3cm}
\paragraph{Eye Tracking Devices.} 
Much like BCI systems, gaze interaction can potentially provide  rich information for the next generation user interfaces.
The most commonly used eye tracking hardware relies on video oculography, capturing videos of user's eyes using a set of cameras. 
Other examples of recent technologies are shown in Figure \ref{fig:eye_gaze_trackers}. 
While some of these prototypes are somewhat bulky and invasive (e.g., Scleral tracker), they will likely influence the design of other unobtrusive eye trackers that might be integrated into thin contact-lens based displays \cite{chen2019design} or yet-to-come technology.
Although the resolution and the sampling rate of a camera are critical for the hardware to offer superior accuracy in eye and gaze location estimations,
recent cameras that have high sampling rates and  resolutions~\cite{angelopoulos2020event} can be demanding in terms of power and computational capabilities. 
These constraints may pose challenges in integrating such hardware with a wearable system such as NEDs~\cite{stengel2015affordable}.

As an alternative to conventional cameras, 
the idea of \textit{event cameras} has been demonstrated as a promising gaze tracking pipeline by capturing only the changing pixels in between frames~\cite{angelopoulos2020event}.
As it can get rid of needs in high-demanding computing power by capturing entire frames every moment (lesser pixels to capture), it could be as fast as 10~kHz.
An interesting direction in eye and gaze tracking relies on using single pixel detectors~\cite{topal2013low,li2020optical, li2017ultra} that may potentially lead to all day useable eye and gaze trackers with low power and low computational complexity.

\begin{figure}[!ht]
    \begin{center}
    \includegraphics[width=\linewidth]{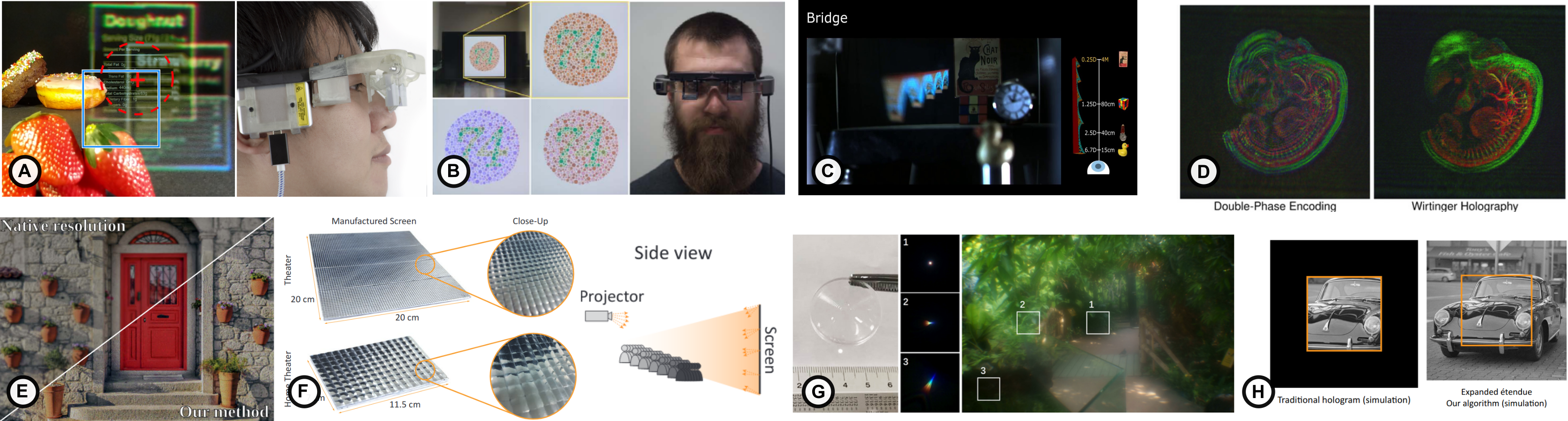}
    \end{center}
    \caption{Some examples of the next generation display technologies discussed in Section~\ref{sec:innovative_hardware}: (a) aoveated near-eye displays~\cite{kim2019foveated}, (b) color correcting near-eye displays~\cite{langlotz2018chroma}, (c) varifocal occlusion capable near-eye displays~\cite{rathinavel2019varifocal}, (d) Wirtinger holography technique for holographic displays~\cite{chakravarthula2019wirtinger}, (e) resolution and refresh rate improvements for next-generation displays~\cite{akcsit2020patch}; (f) directional screens for multiple users~\cite{piovarvci2017directional}, (g) learned Fresnel lens designs for cameras and displays~\cite{peng2019learned}, and (h) improved field of view for holographic displays~\cite{kuo2020high}.}
    \label{fig:display_technologies}
\end{figure}
\vspace{0.3cm}
\subsubsection{Display Technologies.}
In order to facilitate the next generation of telelife technologies, platforms that make use of NEDs~\cite{koulieris2019near}
or 3D displays~\cite{urey2011state} 
will be of the utmost importance.
However, form-factor, high resolution, low latency, and comfortable viewing experiences are still major challenges in adoption for such display hardware.

Like many other hardware design processes, designs of next generation display technologies can be enhanced with the use of machine learning techniques.
In fact, some parts of this transition are already happening with new techniques to design diffractive optical components~\cite{dun2020learned}, fresnel lenses~\cite{peng2019learned}, and to drive holographic displays~\cite{choi2020michelson,peng2020neural}.

Either using machine learning or classical optimization methods, there isn't yet a single design pipeline that optimizes an entire design process of NEDs, while supporting custom needs such as providing prescription support~\cite{chakravarthula2018focusar}, enhanced vision and color capabilities~\cite{langlotz2018chroma}, and opto-mechanical designs that fit perfectly with a user's facial and eye structure.
Beyond these, providing accommodation support in next generation displays regarding heating issues, occlusion support improved optics, dynamic range and field of view are still unresolved points in next generation display designs. 
We also provide examples of futuristic display technologies in the current literature as in Figure~\ref{fig:display_technologies}.

Fortunately, progress is being made in state of the art displays on form-factor~\cite{kuo2020high} and high resolution~\cite{akcsit2020patch}.
In the case of optical see-through displays, it is desirable to be able to change the transmittance of the screen, ideally on a pixel-by-pixel basis \cite{kiyokawa2003occlusion}, and in the case of a video see-through display, it is desirable to be able to see the real environment with imperceptible motion-to-photon latency \cite{lincoln2016motion}.
Many research questions on the transitional interfaces, such as when and how to switch different spaces, and what kind of display hardware is optimal, remain to be answered. 
\vspace{0.5cm}
\subsection{Embodiment}
\subsubsection{Avatars and Agents}
In addition to the technologies that enable telelife, the ways that individuals are represented in virtual space are extremely important. 
Current practice with 2D teleconferencing is to stream 2D video and audio of conversation partners. 
The 3D equivalent is to stream a complete reconstruction of the individual with appropriate occlusion, resolution, lighting, and other visual characteristics~\cite{orts2016holoportation}. 
However, many participants choose to represent themselves using an Avatar, using an image filter, or even as a ghost-camera in the case of certain games. 
Though Avatars are a well-studied field, other possibilities have been relatively unexplored. 

For example, in a 3D online course with 1000 students, it might not make sense to render 1000 individual avatars in a single virtual room. 
Similarly, a teleconferencing window generally can't handle more than a few hundred participants. 
As a new direction in avatar development, avatars should take on more abstract characteristics. 
If a professor needs to view the number of students with questions, perhaps the students should be embodied as a question mark. 
In other words, rendering someone according to their needs or current state may be more useful than rendering a facial expression. 

Alternatively, it might be beneficial for a single simulated physical space to have multiple dimensions.
Imagine a single, 20-person classroom that can hold 1000 students in 50 different dimensions.
The professor is rendered in all dimensions, but students pick one of 50 groups (dimensions) in which to be rendered. 
This dimension division multiple access (DDMA) paradigm may be much more effective for scaling larger groups of people in conferences, classes, or concerts.  

In addition, intelligent agents will also play a key role in the hybridization and scaling of group experiences. 
Much like today's professors utilize teaching assistance to answer questions and grade work, tomorrow's tele-teaching agents could answer questions in real time during class without interruption, facilitate joint note taking between students, and reduce the teaching burden on educators.
Similarly, an agent might assist with a presentation by switching slides, facilitating pointing, or waking sleepy participants.
Individual feedback could also be customized and conveyed by a personal, trusted agent that empathizes with a user, thereby improving the chance that the user will adopt that feedback and improve his or her actions.  

In addition, the geometry of the space in which the user operates in Telelife is generally different from that of the real environment around the user, except when the user is operating in the real environment as is without changing the position or scale, or when the user is operating in a remote or virtual environment with the exact same geometry as the surrounding real environment.
Under such a situation, when the user moves around in the real environment, the geometry mismatch causes the user to collide with the walls or trip over the stuff.
In order to solve this problem, there are some research that provides a virtual fence that is visible but impenetrable, 
or optimizes the relative positional relationship to maximize the range of movement.
While there are many problems caused by the mismatch between real and virtual environments, there have been few studies addressing them.
For example, there is limited research on the use of redirected walking to mitigate the mismatch between real and virtual environments~\cite{sra2018vmotion} in telepresence.
More studies are needed to alleviate the mismatch between real and virtual environments, as it is a key to the success of Telelife.\\
\vspace{0.3cm}
\subsubsection{Multi-modal interactions.}
Multi-modal Interactions are the norm when interacting in the real world but the number of modalities usually decreases during digital interactions.
For example, we communicate not only with our words but also with our bodies and facial expressions. 
While speaking, we can smell things in the environment, feel the movement of air, and sense the temperature among other things.
However, most of these sensory modalities are lost when communicating in a virtual environment and recreating them is challenging, prior work has explored them for specific purposes.
Aspects of conversation beyond speech, like facial expressions and body language are actively under exploration in addition to  new forms of sensory feedback in immersive virtual experiences like temperature and wind~\cite{ranasinghe2017ambiotherm}, 
force feedback~\cite{choi2017grabity,popescu2000virtual},
drag~\cite{jain2016immersive} and weight~\cite{samad2019pseudo}.
There is some work exploring olfaction for PTSD therapy~\cite{rizzo2010development} and gustatory experiences~\cite{narumi2011meta,narumi2011pseudo} 
for attempting to include novel sensations in digital interactions.

However, many challenges in hardware, sensory perception and software need to be overcome before these sensory modalities become commonplace in consumer virtual experiences.
Researchers are starting to explore a new paradigm of human-computer integration that goes beyond human-computer interaction and explores how the muscles in the body and the digital interface form two types of closely coupled systems - symbiosis or fusion ~\cite{mueller2020next}.
This new paradigm presents yet another unique set of challenges and opportunities in the form of technologies that go on, in or through the body, expanding interactions to newer dimensions. 

In addition to audiovisual information, multi-modal presentation is also a major technical challenge.
The receptors for tactile sensation are distributed throughout the body,  
and the senses of smell and taste do not have primary components like the three primary colors of light \cite{miyashita2020}.
To share these senses with the remote environment, multi-modal sensing is required, which poses additional difficulties. 
There is some research on cross-modality, in which one sensory presentation is complemented by another sensory presentation, to generate different flavors through appearance changes \cite{nakano2019}, but the effect is limited.
Multi-modal sensing and display alone are extremely challenging already, but it will be even more challenging to realize multi-modal diminished reality. 
However, the ability to freely modulate these multi-modal information is crucial to enable flexible activities in remote and virtual environments with a high degree of reality, without making the user aware of the physical reality environment.\\
\vspace{0.3cm}
\subsection{State Detection using Biosignals}
In order to develop convincing, valuable, and usable solutions, telelife applications must be able to understand the user's needs, and by extension the user's state, including mental, physical, and emotional aspects. 
In this section, we will look into recent application level research for modalities such as eye tracking, BCIs and sensors that can help identify a user's state. 
Prior work has explored several paths to support users’ cognitive and affective states.
The four most common cognitive states the research community is focused on currently are engagement, attention, cognitive load and fatigue. \\

\subsubsection{Eye movements} 
Eye movements are expressive method for detecting a user's intent, state, or context and they often precede physical action.
Thus, gaze input can support~\cite{ribeletal2013} or substitute~\cite{templieretal2016} manual modes of interaction.
According to Roy et al.~\cite{roy2017novel}, eye fixations can be used to develop a model to predict the objects that a user is observing, which is a hint to how that user is interpreting a scene.
Saccadic eye movements can enable predicting cognitive load, mental fatigue, attention, emotion, anxiety according to work by Duchowski et al.~\cite{duchowski2019using}.
Using eye tracking, the work by Marwecki et al.~\cite{marwecki2019mise} introduces a technique for changing a scene dynamically in VR without being noticed by a user.
In visual search tasks, similar methods (e.g.,~\cite{bektasetal2019},~\cite{gebhardt2019learning}) can be useful for filtering redundant and irrelevant details.

\subsubsection{Engagement} has mostly been studied and tested in learning environments: both subjective and objective methods have been used to provide information to teachers or presenters about the engagement levels of their students or audience ~\cite{10.1145/3025453.3025669}~\cite{s19235200}.
In a recent study titled BrainAtWork~\cite{10.1145/3152832.3152865}, users were presented with their cognitive state, implicitly sensed using electroencephalography (EEG). 
The user’s state was mapped to the workplace activities performed at the time. 
The authors used a visual modality as feedback about the cognitive state of the users, and the study was conducted in a lab. 
Users had the opportunity to see their engagement level both in real time as well as after the experiment was over. 
However, visual feedback might not be ideal for use in cognitively demanding environments such as workplaces, where frequent multitasking between digital windows or tabs within the same application may negatively affect cognitive performance. 
Research projects like AttentivU~\cite{10.1145/3342197.3344516}, a pair of glasses to inform the user about their engagement levels using haptic or auditory modality might be more suitable when designing future AR/VR applications for work/study environments in order not to overload the user's visual field, as well as considering real-time adaptation of the UI based on the physiological state of the user.

\subsubsection{Attention} classification methods are currently not really adapted for workspace use and do not support discrimination between different attention types on a fine-grained level. 
Sohlberg and Mateer’s Clinical Model of Attention~\cite{sohlberg1987effectiveness} discriminates between people’s ability to maintain attention towards a single stimulus (sustained and focused attention); to switch attention between different stimuli (alternating attention); to pay attention to one stimulus while inhibiting others (selective attention), and to pay attention to multiple stimuli simultaneously (divided attention)~\cite{sohlberg1987effectiveness}. 
This model highlights two challenges: quantifying attention (how much attention) and qualifying the nature of attention (what type of attention). 
Prior work on attention has shown that our well-being is tied strongly to our ability to manage attention successfully ~\cite{LEROY2009168}. 
This creates an opportunity to design interactive systems that monitor and actively help users to manage their attention.
The vision of pervasive attentive user interfaces encapsulates this well, stating that interfaces could “adapt the amount and type of information based on users’ current attention capacity, thereby simultaneously optimizing for information throughput and subtlety. 
Digital interfaces of the future could trade-off information importance with users’ current interruptibility level and time the delivery of information appropriately”~\cite{10.1109/MC.2016.32}. 

\subsubsection{Cognitive load} has been measured traditionally either by standard questionnaires or by measuring user's task performance. 
The NASA TLX is a common example of the first approach, where participants are asked to report their own cognitive load with regard to six different categories. 
Another example where study participants are asked to report their own estimates can be found in Sweller et al. ~\cite{Sweller2011}. 
One drawback of these approaches is that the answers are subjective.
Furthermore, the self-reporting itself adds to the cognitive load.
Measuring cognitive load through the performance in the task itself or in a secondary task (e.g. Lane Change Task for Automotive user interface, ISO 26022) only provides a rough estimate and is typically only suitable for laboratory studies and not for creating cognition-aware real-time systems. 
For interactive systems to be able to adapt their behavior accordingly, cognitive load information must be captured continuously and automatically—introspection is often not sufficient.
Physiological sensors, such as EEG, and electrodermal activity (EDA) sensors show potential as possible solutions to this problem.\\ 

\subsubsection{Fatigue} 
Fatigue can be defined as the unwillingness to continue performance of mental work in alert and motivated participants~\cite{MONTGOMERY1995143}. It affects different cognitive functions including alertness, working memory, long-term memory recall, situational awareness, judgment, and executive control.
Several technologies exist for monitoring fatigue levels, including eye tracking~\cite{10.1145/3079628.3079669} as well as video recording~\cite{Raca2015TranslatingHM}. 
These solutions are prone to errors and have limitations as cameras are sensitive to the ambient illumination and pose privacy problems, and are constrained to specific locations. 
Other measures include sensing physiological signals such as heart-rate variability ~\cite{10.1007/978-3-540-73331-7_70}, electrodermal activity ~\cite{BYRNE1996249}, brain activity signals like EEG~\cite{Zander2010EnhancingHI}, Electromyography (EMG) ~\cite{Fu2016DynamicDF},and Electrooculography (EOG)~\cite{8771080}. 

Current knowledge on the potential use of several types of sensors to measure each of these four phenomena in real-time on a fine-grained level is summarized in Figure \ref{fig:bci_devices}.
Despite multiple recent publications which investigate attention, cognitive load, fatigue and engagement, very few approaches have actually been deployed in the real world.
The challenges to make this happen include the choice of modality or modalities with which to precisely measure one or more phenomena, hardware limitations (social acceptability of the device(s), form-factor, and comfort), and user value of these tools. Currently, passive sensing is only offered to the user, with very few options for adaptation of their environment and/or other active interventions.\\

\subsection{New Applications}
The research directions mentioned above will ultimately be driven by a set of applications within Telelife that call for the advance of technology. 
We present several of those applications below that will be at the forefront of remote interactions. \\

\subsubsection{Gaze-Contingent and Context-Aware Assistance}  
Contextual information characterizes the interaction of a user with a computational system~\cite{deyabowd2000}.
Context-aware systems provide information and services that are relevant to the current activity or task of the \textit{user} and they are expected to adapt to dynamically changing situations of the inhabited \textit{environment} and the \textit{system} itself.
According to Grubert et al.~\cite{grubertetal2017} these three factors constitute the context sources.
In a meeting context, audio and visual inputs to a computer can be disabled and tactile input (e.g., typing) can be enabled instead.
In a training context, task-relevant visual content can be highlighted if it remains beyond the user’s attention.
An assessment of user and system behavior (preferably in real time) is a prerequisite to these adaptations.

One way of understanding user behavior is to assess their visual perception through gaze-enabled systems.
Eye tracking has recently become a promising feature in many HMDs because it enables foveated rendering, depth-of-field simulation, and studying the user’s viewing behavior in the context of collaborative VR~\cite{stengel2016} and AR applications~\cite{kim2019foveated}.
By mimicking human vision, gaze-enabled systems can substantially improve rendering ~\cite{kim2019foveated, bektacs2015testbed} and user performance \cite{bektasetal2019}. 
Such systems can direct user attention, augment their vision with contextually relevant information, and provide personalized assistance in different activities~\cite{bektas2020toward, gardonyetal2020}.
Some future implications of remotely performed activities may be beyond comprehension, so understanding how physical things and their digital counterparts inhabit telelife will play a major role in future research \cite{mayer2018hypermedia}.
\vspace{0.3cm}
\subsubsection{Digital Twins and Digital Companions}

\begin{figure}[!ht]
    \begin{center}
    \includegraphics[width=0.48\textwidth]{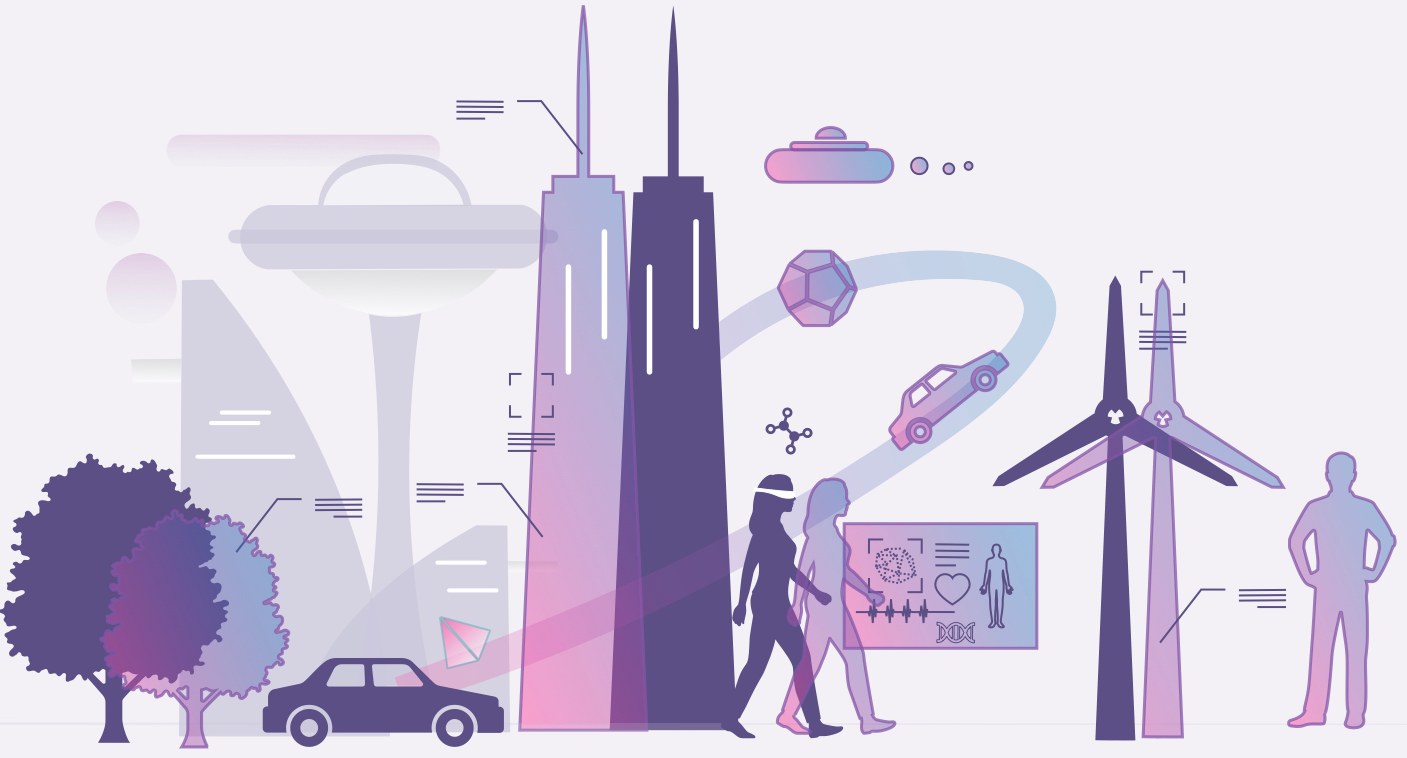}
    \end{center}
    \caption{Conceptual image highlighting a world in which the digital twins of people, buildings, vehicles, nature, etc. are integrated into the fabric of daily life.}
    \label{fig:digital-twins}
\end{figure}

Our future tends to be a hybrid one that will be composed of smart environments where information from physical, social, and digital sources is intertwined~\cite{ricciMirror2015}.
The digital twin of a physical system accounts for its real-world interactions between the environment and physical models through realistic simulations to update changes accordingly \cite{glaessgen2012digital}. 
The twin is a living model that continuously adapts operational changes based on the collected data and simulation, either in real time or accumulated, and would allow forecasting the future of the corresponding physical counterpart~\cite{liuetal2018}.
The twin will thus enable \textit{mirroring} the life of its physical counterpart~\cite{glaessgen2012digital}.
In our telelife vision, digital and physical beings (including humans) will grow with and adapt to each other. 
This adaptation needs to be carefully studied as it has the potential for expanding our perceptual capabilities while depriving others.
To this end, monitoring behavioral changes will allow us to understand how users' cognitive abilities can be augmented in an environment with a digital twin. 

Digital companions are informed by semantic representations of context sources~\cite{garciaetal2018}, and can provide autonomous assistance to users by proactively acting, sensing, tracking, and communicating on their behalf.
Recent work predicts that the decision support using digital twins will feed into these companions, which will be essential for industrial applications~\cite{hartmann2020digital} and possibly part of our daily life by 2030.

Despite some practical challenges, mobile and wearable sensors allow us to develop BCIs that are dedicated to monitoring a user's cognitive state based on psycho-physical measures.
Such interfaces hold the promise to support operators, maintenance workers, and other users by eliciting device interactions in pervasive computing environments.
In the near future, the citizens of telelife may proactively remain at the center of the computation-loop~\cite{shneiderman2020}, while advanced MR interfaces support human-machine interactions and provide customized assistance.
\newpage
\subsubsection{Virtual Prototyping and Remote Fabrication} 
Telelife will also benefit from advancements in other tools and systems that can physically produce and share virtual designs. 
An existing example is that of construction robots, which can build artifacts at remote locations where it would not be possible for a human supervisor to be safely present\footnote{ \url{https://www.therobotreport.com/construction-robotics-changing-industry/}}.
We have already seen a shift towards remote fabrication, with access to personal fabrication tools (e.g., laser cutters, 3D printers) showing it is possible for open-source custom manufacturing in everyday life. 
Besides the growing number of consumer-grade fabrication tools and virtual resources, the functionality of software and hardware systems is also improving. 
Several advances~\cite{peng2016fly, kelly2019volumetric} support physical instantiation at a much faster speeds.
The progress of these technologies suggests a future where household hardware can be customized and created in-situ. This also highlights the need for education in design and fabrication skills, which will be a significant component of telelife.

Much recent research has made progress in enabling possibilities, such as RoMA~\cite{Peng2018RoMA}, a system that combines 3D digital modeling with augmented reality and 3D printing. This supports seamless 3D construction by allowing a user to directly design in AR and print in real-time. 
In a similar context, compositional 3D printing~\cite{Kim2018Compositional} further promotes embodied fabrication using multi-modal input to support seamless construction of physical models during design.
Scotty~\cite{Mueller2015Scotty}, a self-contained appliance that allows relocating inanimate physical objects across distance, is another example of how remote physical editing will play a role in Telelife. 
The user can send a physical copy to a remote place with the local one being destroyed, emulating the concept of teleportation, while at the same time demonstrating a creative way to protect copyrights. 
Scotty can be further equipped with AR editing features in which the teleported artifact can also be custom designed on-the-fly.

This body of work will be essential for remote collaboration.
For example, during a digital meeting across the globe, one can design artifacts such as personal garments or large architecture at remote workshops. While the design could be fabricated locally by receiving real-time data, a consultant would have the opportunity to participate in the process, seamlessly updating custom requirements and other real-world constraints, visualized in the local space with a digital twin.
\section{Grand technical challenges}
Existing platforms for online ecosystems are in many ways well established, but they are still generally divided into separate fields. 
Here, we aggregate the most important challenges from these fields into unified grand challenges that represent the most important problems for the future of remote living. 
This is also designed to give researchers a holistic picture of the problem space for each area in order to better guide current and future research. 

\subsection{Re-humanized teleinteractions.} 
Tomorrow's technologies have the power to replicate the social and emotional interactions that we have lost in today's 2D world of social media. 
As such, we need to focus on re-enabling the face-to-face and interpersonal interactions that define human contact. 
The technology will not only have to be closely aligned with human nature, but will need to be accurate enough to reproduce facial expressions and body language.
The grand challenge here is to reverse the process caused by 2D interfaces that has slowly detracted from interpersonal relationships and to re-humanize interaction so that remote living meets or exceeds the benefits of face-to-face contact. 

\subsection{Perfect telepresence.} An experience in which the physical and virtual are indiscernible from each other will require a fully integrated set of new hardware devices.
These devices must render perceptually realistic experiences that support all senses (visual, auditory, tactile, smell, taste).
To this end, true 3D displays will help unlock perfect visual experiences.
Shape changing haptics will enable touch-experiences to render and augment physical objects of any shape or physical space.
The combination with olfactory, auditory, and gustatory modalities will help reproduce experiences that can provide a true one-to-one correspondence to a user's physical body. 
The challenge with perfect telepresence is to address all of these sensory areas in a comprehensive display system. \\

\subsection{Complete cognitive sensing.}
Understanding a user's needs, including emotional, social, intellectual, and innate, will enable interfaces that can better care for and facilitate communication in future societies.
Two major research areas that show promise for the understanding of cognitive and contextual state include brain-computer and gaze tracking.
Future components must be unobtrusive, support all-day usability, have seamless integration into eye- or head-wear, and require low computational and power resources.
Artificial intelligence and machine learning will be indispensable in this task, where the main challenge is to provide consistent, accurate estimates of the broad range of a user's cognitive states.\\

\subsection{Contextual teleinterfaces.}
Our lives will be changed by future technologies, and our societies have to find an effective and beneficial way to cope with such changes per individual and for all members of society.
From the design perspective, diverse virtual spaces and transitional interfaces \cite{grasset2006} that continuously provide a diversity in space types throughout a day need to be explored in more depth to switch smoothly from one task to another while maintaining cognitive continuity and physical safety, and not overwhelming users with monotonous experiences.
Designs of these new experiences has to properly inform a user on what is virtual and what is part of real life, so that users can cope with the difference between virtual and real.\\

\subsection{Teleaccess for all.} A current major bottleneck in delivering online services is the lack of computing resources needed to provide such services to every member of society.
Either in a local or cloud context, a wide variety of telelife experiences must be supported with features that meet minimum standards of teleliving.
Zero social isolation is a critical component in achieving this goal. 
The grand challenge here is to provide access, define minimum standards of teleliving, and design sustainable resources such that all individuals can live a full telelife. 

In summary, Telelife should be universally accessible, give us the best experience with the right set of technologies, understand our emotional and social needs, provide us with an interface that supports mental health and growth, and mitigate potential dangers to our wellbeing to the greatest extent possible.

\section{Conclusion}
The rapid evolution of technology has brought new opportunities for all of us.
However, the pace of this evolution does not necessarily provide enough time to build a roadmap for the future ahead and seek means to pursue ethical designs and processes.
Laying a high-level tangible roadmap can provide the opportunity to effectively solve problems by uniting research communities to move towards a common goal.
We believe the current communities involved in virtual and augmented reality need such a roadmap to impact the way our societies live positively.

Therefore, we ask ourselves a simple question: How should the remote technologies of the future look like, and how can we ensure that the technologies remain well aligned with humanly needs. 
Our answer to this question is the vision that we call telelife, a high-level term that encapsulates previously introduced visions and concepts such as telepresence, telework and AR/VR. 
We cherry-pick and merge the essential items from those fields, identify relevant challenges in achieving telelife, provide an overview of the state-of-the-art, and outline future directions in research that will help re-humanize remote technologies and teleinteraction.

Remote living is a multi-dimensional problem larger than all of us as individuals and will require efforts from each of our unique research fields. We hope to unify the visions of researchers so that their work will better fit into the upcoming telelife ecosystem. 
Moreover, the concepts presented here can inspire and guide future generations of research on remote living.
Lastly, we aim to raise the opportunity to discuss future questions about how we can and \textit{should} live our remote social, intellectual, professional, and personal lives.

\section*{Funding}
This work was funded in part by the University College London and Osaka University partnership fund, grant \#Na20990020, and by the Office Of Naval Research Global, grant \#N62909-18-1-2036.

\section*{Acknowledgments}
The authors would like to thank reviewers for their valuable feedback. Special thanks to Michael Nowatkowski from Augusta University for his input. 
\bibliographystyle{frontiersinSCNS_ENG_HUMS}
\bibliography{references}

\end{document}